\documentclass{egpubl}
\usepackage{sca2022}
 
\ConferenceSubmission   % uncomment for Conference submission
\ConferencePaper        % uncomment for (final) Conference Paper
\SpecialIssuePaper         % uncomment for final version of Computer Graphics Forum, special issue
\usepackage[T1]{fontenc}
\usepackage{dfadobe}  

\biberVersion
\BibtexOrBiblatex
\usepackage[backend=biber,bibstyle=EG,citestyle=alphabetic,backref=true]{biblatex} 
\electronicVersion
\PrintedOrElectronic

\ifpdf \usepackage[pdftex]{graphicx} \pdfcompresslevel=9
\else \usepackage[dvips]{graphicx} \fi

\usepackage{egweblnk} 
\title{Wassersplines for Neural Vector Field--Controlled Animation}
\author[P. Zhang \& D. Smirnov \& J. Solomon]
{\parbox{\textwidth}{\centering P. Zhang$^{1}$
        and 
        D. Smirnov$^{1}$ 
        and
        J. Solomon$^{1}$
        }
        \\
{\parbox{\textwidth}{\centering $^1$Massachusetts Institute of Technology, USA}}}
\makeatletter

\newcommand{\R}{\mathbb{R}}

\newcommand{\Tr}{\text{Tr}}
\newcommand{\K}{\mathcal{K}}
\newcommand{\X}{\mathcal{X}}

\newcommand{\lfit}{\text{fit}}
\newcommand{\ldiv}{\text{div}}
\newcommand{\lrigid}{\text{rig}}
\newcommand{\lcurl}{\text{curl}}

\newcommand{\lvel}{\text{vel}}
\newcommand{\laccel}{\text{acc}}
\newcommand{\ltotal}{\text{tot}}
\newcommand{\ljerk}{\text{jerk}}
\newcommand{\lgrad}{\text{grad}}
\newcommand{\lmetric}{\text{A}}
\newcommand{\OT}{\text{OT}}
\newcommand{\KL}{\text{KL}}

\newcommand{\argmin}{\text{argmin}}

\newcommand{\citet}[1]{\cite{#1}}
\newcommand{\citep}[1]{\cite{#1}}
\usepackage{lipsum}
\usepackage{nicefrac}
\usepackage{placeins}
\usepackage{float}
\usepackage{amsfonts}
\usepackage{amsmath}
\renewcommand{\sectionautorefname}{\S\@gobble}
\makeatother

\newcommand{\add}[1]{{#1}}
\newcommand{\deletesec}[1]{{}}
\DeclareRobustCommand{\delete}[1]{{}}

\begin{document}

\teaser{
 \includegraphics[width=\textwidth]{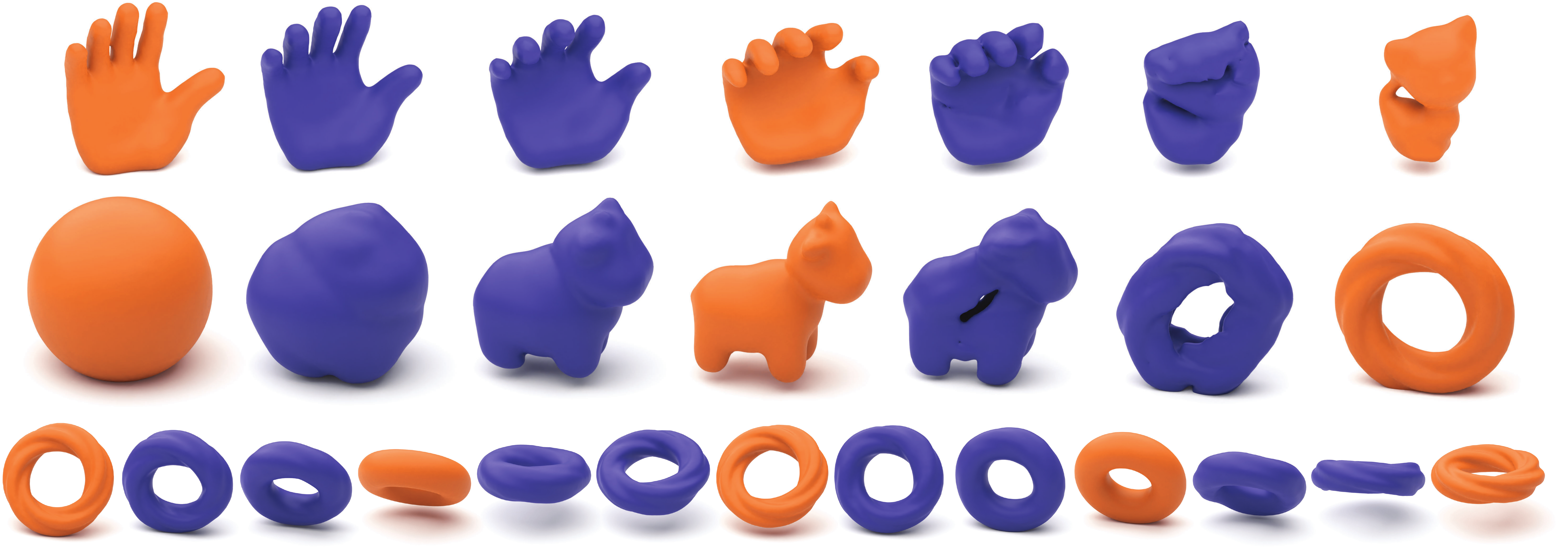}
 \centering
  \caption{We apply our method on 3D data to generate smooth animations between keyframes (orange) of varied geometry and topology.}
    \label{fig:3d-traj}
}
\maketitle

\begin{abstract}
Much of computer-generated animation is created by manipulating meshes with rigs. While this approach works well for animating articulated objects like animals, it has limited flexibility for animating less structured free-form objects.  We introduce Wassersplines, a novel trajectory inference method for animating unstructured densities based on recent advances in continuous normalizing flows and optimal transport. The key idea is to train a neurally-parameterized velocity field that represents the motion between keyframes. Trajectories are then computed by advecting keyframes through the velocity field. We solve an additional Wasserstein barycenter interpolation problem to guarantee strict adherence to keyframes. Our tool can stylize trajectories through a variety of PDE-based regularizers to create different visual effects. We demonstrate our tool on various keyframe interpolation problems to produce temporally-coherent animations without meshing or rigging.


\end{abstract}

% \begin{teaserfigure}
%     \centering
%     \vspace{-5pt}
%     \includegraphics[width=\textwidth]{figures/3d_traj}
%     \vspace{-22pt}
%     \caption{We apply our method on 3D data to generate smooth animations between keyframes (shown in orange) of varied geometry and topology.}
%     \label{fig:3d-traj}
% \end{teaserfigure}

\maketitle
\section{Introduction}

In hand-drawn animation, a primary artist is tasked with laying out \emph{keyframes}. These frames define the rough motion of an animation and occupy a fraction of the usual 12 drawings per second. The remaining frames are filled in afterwards to create smooth motion in a process called \emph{inbetweening}. In the transition from hand-drawn animation to computer-assisted animation, much of the inbetweening process became automated \cite{lasseter1987principles}. After an artist lays out keyframes as mesh rig displacements, in-between frames can be produced automatically using splines and other interpolation machinery. % by interpolating rig displacements between keyframes. 
Secondary effects like elastic oscillation or fluids can be added afterwards through \add{physical simulation \cite{kass2008animating}}. This process is largely responsible for modern character animation and has had major success for articulated objects like humans and animals.

While methods for articulated animation via meshes and rigs are abundant, research on animation of objects like the Drunn in Walt Disney Animation Studio's ``Raya and the Last Dragon'' is far less common. These animations are characterized by abstract, amorphous boundaries and free-form motion. As with classical animation, keyframes are still provided by an artist to coarsely define the desired motion. 
% Contrary to rig-based methods, however, the object being animated has no well-defined skeletal structure around which to build a rig. 
Due to the lack of structure between keyframes, however, we denote such animations as \emph{unstructured}. 
Rig-based methods are insufficient in this case, since unstructured animations can separate and recombine. Simultaneous to their freedom of movement, their trajectories must accurately reach keyframes, so that an artist is able to convey the appropriate gestures in a scene.

% Animating the Drunn required interpolation of point clouds embedded in Drunn's amorphous bodies; optimal transport methods were applied to reduce the necessary manual effort \cite{drunn}. 
%, they developed an automatic technique by concatenating optimal transport maps

Unstructured animation can be found in various media over the last several decades.
% In 2018, Supergiant's ``Hades'' animated trailer depicts a hand drawn sequence of smoke assembling into the form of Zeus.
% In 1994, Nickelodeon's ``The Secret World of Alex Mack'' depicts Alex dissolving into a fluid and reforming. 
In 1991, ``Terminator 2: Judgment Day'' depicts visual effects of mercury transitioning into various geometries. 
The unstructured animation style even pre-dates computer animation and can be found in the exaggerated motions of Cruella's cigarette smoke in Disney's 1961 ``One Hundred and One Dalmations.''
That is to say, interest in this type of animation is abundant, but methods are largely manual, highly specific to the scene, and undocumented. 

% drunn 2021
% umbrella academy - 2019
% spiderman far from home - 2019
% hades - game 2018
% moana 2016
% animorphs - 1998
% the secret world of alex mack - 1994
% terminator 2: judgement day. 1991 
% In the season 2 opening of Netflix's ``Umbrella Academy`` a flock of birds drift into an umbrella shaped keyframe before scattering again. 

In this paper, we present \emph{Wassersplines} for unstructured animation. We encode keyframes as point clouds or probability measures, allowing us to capture arbitrary geometries without the limitations of a mesh or rig. Trajectories are encoded using a coordinate multi-layer perceptron (MLP) to produce a velocity field in space-time. A rough animation is then produced by advecting points through the velocity field. We propose a Wasserstein barycenter interpolation step to guarantee strict keyframe adherence. Using partial differential equation-- (PDE--) based regularizers on the coordinate MLP, we bring about various effects in the animation without needing extra keyframes. We demonstrate our method on 2D and 3D examples, showing strict adherence to keyframes and flexibility in the interpolations.

\vspace{-5pt}
\section{Related Work}

\paragraph*{Normalizing Flows.} Normalizing flows map a prescribed initial density, such as a Gaussian, through a set of invertible functions to a target posterior distribution \cite{rezende2015variational,papamakarios2019normalizing}. Neural ordinary differential equations (ODE) take this concept to the limit by parameterizing the state derivative with a deep neural network. Continuous normalizing flows (CNF) integrate the neural ODE to produce the target posterior distribution \cite{neuralode}. In this context, \citet{ffjord} use the Hutchinson's trace estimator to compute posterior density values through a neural ODE.

Various strategies decrease training time for CNFs. \citet{kelly2020learning, finlay2020train} regularize the spatial variation of the state derivative and its higher-order time derivatives. \citet{tancik2020fourier} use random Fourier features (RFF) for faster training of high-frequency state derivatives. \citet{hertz2021sape} sequentially unmask RFFs in order of increasing frequency, decreasing sensitivity to the initial RFF sampling. \citet{poli2020hypersolvers} learn auxiliary networks for faster ODE integration, a costly step in application of CNFs. 

\paragraph*{Optimal Transport.} Optimal transport (OT) models compute the cheapest map from a source distribution onto a target distribution via a linear program \cite{kantorovich2006translocation}. The cost of this map defines the \emph{Wasserstein distance} between distributions; see \cite{peyre2019computational, solomon2018optimal, santambrogio2015optimal} for general discussion.

Adding entropic regularization to the optimal transport linear program yields an efficient and easily-implemented optimization technique known as Sinkhorn's algorithm or matrix rebalancing \cite{cuturi13}. While cheaper to compute, entropically-regularized optimal transport biases the Wasserstein metric so that the distance from a distribution to itself is nonzero. This issue is fixed in the definition of the \emph{Sinkhorn divergence} by adding a de-biasing term to entropic optimal transport \cite{genevay2018learning}. Efficient large-scale implementations of Sinkhorn divergence and other optimal transport routines are available through ``GeomLoss'' \cite{feydy2019interpolating} and ``POT: Python Optimal Transport'' \cite{flamary2021pot}.

\emph{Dynamical} optimal transport provides an alternative means of computing Wasserstein distances when the ground metric is quadratic in geodesic distance. Instead of computing a map between the source and target distributions, it computes a kinetic energy-minimizing velocity field that advects the source distribution into the target \cite{benamou2000computational}; recent algorithms accelerate solution of the relevant variational problems and explore alternative mesh-based and neural parameterizations \cite{lavenant2018dynamical, lavenant2021unconditional, papadakis2014optimal, trajectorynet}. 

\paragraph*{Image/Shape Registration.}
Registrations between images or shapes can be built from velocity field--induced diffeomorphisms. \citet{hug2015multi} regularize the velocity field in dynamical optimal transport and prove existence of minimizers with a velocity gradient regularizer. % apply div free reg to ocean height maps
\citet{eisenberger2018divergence} compute static volume-preserving, velocity fields for mesh registration.
\citet{feydy2017optimal} use unbalanced OT to build diffeomorphic registrations in medical imaging. 
% lddmm - large def diffeomorphic metric mapping. 
% Rather than using velocity fields to compute registrations, \citet{solomon2019optimal} use registrations between vector fields singularities to compute intermediate vector fields.

\paragraph*{Measure-valued Splines.}
Higher-order interpolations can be computed through an ordered sequence of distributions by minimizing acceleration instead of kinetic energy \cite{chen2018measure,benamou2019second}. \citet{benamou2019second} compute solutions as distributions over cubic splines via a multi-marginal transport problem. % with Sinkhorn's algorithm. 
\citet{chewi2021fast} show that such solutions do not allow for deterministic trajectory inference and instead %propose the following alternative. Assuming distributions are represented by point cloud samples of the same size, \citet{chewi2021fast} 
compute optimal transport plans between consecutive pairs of point clouds followed by classical spline interpolation.

Our problem also aims to interpolate an input sequence of distributions. A major difference, however, is that we 
parameterize the trajectory of our interpolation with a velocity field. This difference allows us to regularize based on spatial derivatives of the velocity rather than just time-derivatives like acceleration. 
Furthermore, our aim is not to globally minimize any particular time derivative but rather to provide a palette of effects to control a trajectory.
%Furthermore, the use of a smooth velocity field gives our solutions preference to trajectories that do not split mass. 

\paragraph*{Trajectory Inference.}
\citet{schiebinger2019optimal, lavenant2021towards} propose 
\emph{Waddington OT} for inference of cellular dynamics by concatenating OT interpolations between consecutive keyframes. 
%Waddington OT produces non-smooth trajectories since each keyframe only influences its immediately-adjacent trajectories.
The approach used by \citet{drunn} to animate Drunn is similar in that it also computes OT matchings between consecutive point clouds. A key difference, however, is that the point clouds used in animation of Drunn are automatically generated by sampling densities evolved via fluid simulation. This provides an abundance of data, mitigating artifacts of the piecewise-smooth trajectory. 
\citet{tang2021honey, treuille2003keyframe, pan2017efficient} augment incompressible fluid simulations with additional forces to interpolate between target keyframes. These methods balance between proximity to keyframes and regularization of additional forces thus allowing deviation from target keyframes. 
% \add{\citet{kass2008animating} augment rig based interpolation with physics-based oscillations to produce more realistic motions.}
\citet{browning2014stylized}
interpolate keyframes by matching image patches between keyframes. To get smoother trajectories, however, they also allow deviation from the provided keyframes. 

% Most similar to our work is TrajectoryNet \cite{trajectorynet}. Given keyframe data in the form of point clouds, they compute continuous normalizing flows that minimize the KL divergence from the mapped density to target keyframe as well as kinetic energy of the trajectory. They show that for two keyframes, given a large enough weight on the KL divergence term, their solution trajectory is the same as one computed with dynamical optimal transport. Their method is applied towards inference of cellular dynamics, where exact keyframe fitting is not required and cell distributions are assumed to have infinite support. Since both of these assumptions are unsuitable for animation, our method ultimately deviates from theirs despite having similar usage of point cloud data and continuous normalizing flows.

\section{Preliminaries}
\label{sec:prelims}

For completeness, we overview relevant developments in continuous normalizing flows and optimal transport. For detailed coverage, see \cite{papamakarios2019normalizing,peyre2019computational,genevay2018learning}.

\subsection{Neural ODEs and CNFs}
CNFs fit a target measure as the pushforward of a source measure through a diffeomorphism, granting users the ability to sample the target measure as long as they have sample access to the source measure. We will re-purpose tools from the CNF literature to generate unstructured animations and review their basics here.

Given an initial state $z(t_0)\in\R^n$ and parameterized state derivative function $f_\theta(z,t)\in\R^n$, one can solve the ODE $\dot{z} = f_\theta(z,t)$ for $z$ at time $t_1$ as
\begin{equation}
    z(t_1) = z(t_0) + \int_{t_0}^{t_1} f_\theta(z(t),t) dt.
    \label{eq:ode}
\end{equation}
When $f_\theta(z,t)$ is parameterized by a deep neural network, it is referred to as a \emph{coordinate MLP}; $\dot{z} = f_\theta(z,t)$ is a \emph{neural ODE}. Given a loss function $L(z(t_1))$, the gradient $\nabla_\theta L(z(t_1))$ is computed by the adjoint method using black box ODE solvers \cite{neuralode}.

Neural ODEs build CNFs in the following way. Let $\nu$ be a source measure with simple parametric density on $\R^n$, $\mu$ be the measure of a target density, and $\phi$ be a map from $z(t_0)$ to $z(t_1)$. Then $\phi_{\#}\nu$ denotes the \emph{pushforward measure} obtained by advecting $\nu$ through $f_\theta$ from $t_0$ to $t_1$, i.e., the pushforward of $\nu$ by $\phi$. The CNF generating $\mu$ is obtained by minimizing Kullback-Leibler (KL) divergence $\KL(\mu | \phi_\# \nu)$ with
\begin{equation}
    \KL(p | q) = \int_{\R^n} \log\left(\frac{dp}{dq}\right) dp
\end{equation}
for measures $p$, $q$ over $\R^n$ \cite{papamakarios2019normalizing,neuralode}.

\subsection{From Wasserstein Distance to Sinkhorn Divergence}
While KL divergence is a popular loss function for building CNFs, it requires overlapping support and density access to work. To bypass these shortcomings, we use Sinkhorn Divergence instead and introduce it here.

Given probability measures $\mu$ and $\nu$ on $\X \subset \R^n$, let $\Pi(\mu,\nu)$ denote the set of joint probability measures on $\X\times\X$ with marginals $\mu$ and $\nu$. The squared \emph{2-Wasserstein distance} between $\mu$ and $\nu$ is defined as
\begin{equation}
    W_2^2(\mu,\nu) := \inf_{\pi\in\Pi(\mu,\nu)} 
    \int_{\X \times \X}
    \|x-y\|^2 d\pi(x,y).
    \label{eq:wass2}
\end{equation}
The joint probability measure solving \autoref{eq:wass2} is the optimal transport plan $\pi$. The 2-Wasserstein distance is well defined even when $\mu$ and $\nu$ are non-overlapping, and its gradient brings non-overlapping measures together.

\autoref{eq:wass2} is computationally challenging to solve, leading to the popular use of entropic regularization. Entropically-regularized transport distance is defined via the convex program
\begin{equation}
    \OT_{\epsilon}(\mu,\nu) := \inf_{\pi\in\Pi(\mu,\nu)} 
    \int_{\X \times \X}\|x-y\|^2 d\pi(x,y) + 
    \epsilon KL(\pi \mid \mu \otimes \nu).
    \label{eq:entropicOT}
\end{equation}
Unlike \autoref{eq:wass2}, $\OT_\epsilon(\mu,\nu)$ can be computed efficiently by Sinkhorn's algorithm \cite{cuturi13}. 
This efficiency comes at the cost of entropic bias, i.e., $\OT_\epsilon(\mu,\mu)\neq0$. The bias becomes especially problematic when one is interested in Wasserstein gradient flows to transform a source measure $\nu$ into a target measure $\mu$. One could implement the flow $\dot{\nu} = -\nabla_\nu OT_\epsilon(\mu,\nu)$, but it would converge to a solution where $\nu\neq\mu$ \cite{feydy2019interpolating}. To address this bias, \citet{genevay2018learning} build the \emph{Sinkhorn divergence}:
\begin{equation}
    S_{\epsilon}(\mu,\nu) := \OT_\epsilon(\mu,\nu) 
    - \frac{1}{2}\OT_\epsilon(\mu,\mu) 
    - \frac{1}{2}\OT_\epsilon(\nu,\nu).
    \label{eq:sinkdiv}
\end{equation}
$S_{\epsilon}(\mu,\nu)$ eliminates entropic bias and restores the desired property $S_{\epsilon}(\mu,\mu)=0$. 
When $\epsilon=0$, we have the equivalence $$S_0(\mu,\nu)=\OT_0(\mu,\nu)=W_2^2(\mu,\nu),$$ but the computational benefits of entropic transport are lost.

\subsection{Unbalanced Optimal Transport}
\label{sec:prelims_unbalanced}
\emph{Unbalanced optimal transport} is an extension of OT where marginal constraints are softened and controlled by an extra parameter $\tau$. The softened constraints make unbalanced OT more robust to outliers, which we use in 
% \autoref{fig:balanced_hand}
\autoref{sec:results}
to improve quality of results.
Let $\mathcal{M}^+(\X\times\X)$ be the space of positive measures on $\X\times\X$. Then the unbalanced OT cost is
\begin{align}
    \OT_{\epsilon,\tau}(\mu,\nu) & :=  \inf_{\pi\in\mathcal{M}^+(\X\times\X)} 
    \int_{\X \times \X}\|x-y\|^2 d\pi(x,y)  \nonumber
    \\
    & + \epsilon KL(\pi \mid \mu \otimes \nu) + \tau^2 KL(\pi_1 \mid \mu) + \tau^2 KL(\pi_2 \mid \nu),
    \label{eq:unbalancedentropicOT}
\end{align}
where $\pi_1$ and $\pi_2$ are marginals of $\pi$ \cite{feydy2020geometric}, and $\tau$  can be intuitively thought of as the maximum distance a piece of mass can be transported before the transport plan $\pi$ would rather violate marginal constraints. $\tau$ is referred to as ``reach'' in GeomLoss \cite{feydy2017optimal,feydy2019interpolating}.

Unbalanced Sinkhorn divergence  $S_{\epsilon,\tau}$ is defined exactly the same as $S_{\epsilon}$ in \autoref{eq:sinkdiv} but with $\OT_\epsilon$ replaced by $\OT_{\epsilon,\tau}$. 
Despite the softening of marginal constraints, $S_{\epsilon,\tau}(\mu,\nu)=0$ still implies $\mu=\nu$. When, $\tau=\infty$, $S_{\epsilon,\tau}=S_{\epsilon}$ i.e. balanced Sinkhorn divergence.

\section{Method}
We now present unstructured animation as a measure interpolation problem equipped with a suitable fitting loss. We describe how to control the animation through different regularizers as well as how to guarantee strict adherence to keyframes.

\begin{figure}
    \centering
    \includegraphics[draft=false]{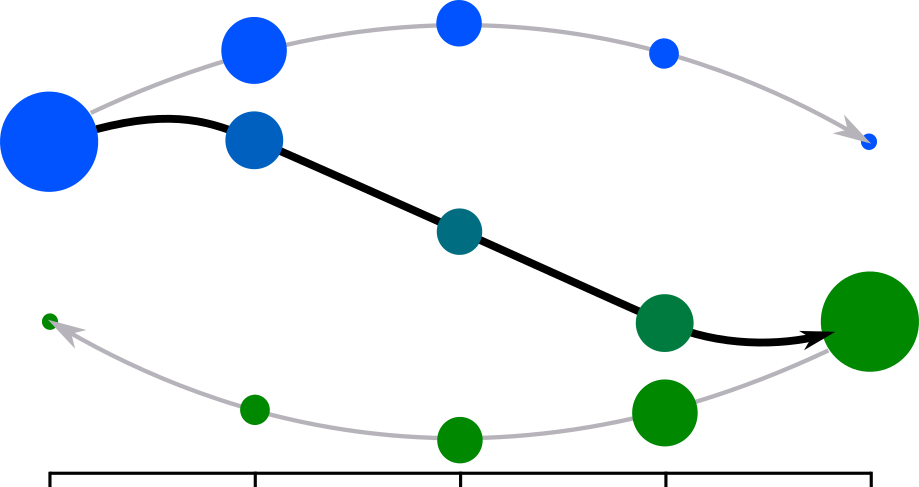}
    \put(-215,100){$\X_i$}
    \put(-115,122){$\X^t_i$}
    \put(-115,22){$\X^t_{i+1}$}
    \put(-20,95){$\X^{t_{i+1}}_i$}
    \put(-215,25){$\X^{t_i}_{i+1}$}
    \put(-20,20){$\X_{i+1}$}
    \put(-115,70){$\X(t)$}
    \vspace{-10pt}
    \caption{Schematic of Wasserstein barycenter interpolation. While the neural ODE--produced trajectories (in grey) do not precisely adhere to keyframes, the modified trajectory produced using \autoref{eq:wassinterp} (in black) is guaranteed to adhere to keyframes.}
    \label{fig:wassinterp}
\end{figure}

\begin{figure}
    \centering
    \includegraphics[width=\columnwidth]{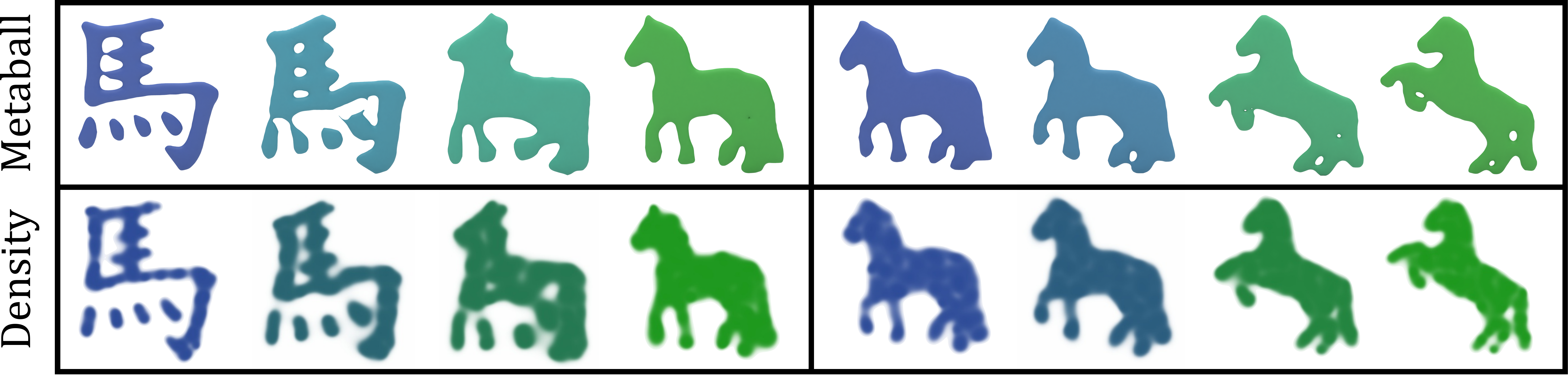}
    \vspace{-15pt}
    \caption{Different renderings of our results. The first row is rendered by converting point clouds into an explicit surface using metaballs with a 0.05 radius. This technique results in more well defined boundaries but less visible interior density variation. The second row is rendered directly as densities computed via \cite{bonneel2011displacement}. Boundaries are less sharp. but more interior variation is revealed. }
    \label{fig:doublerender}
\end{figure}

\begin{figure*}
    \centering
    \includegraphics[width=\textwidth]{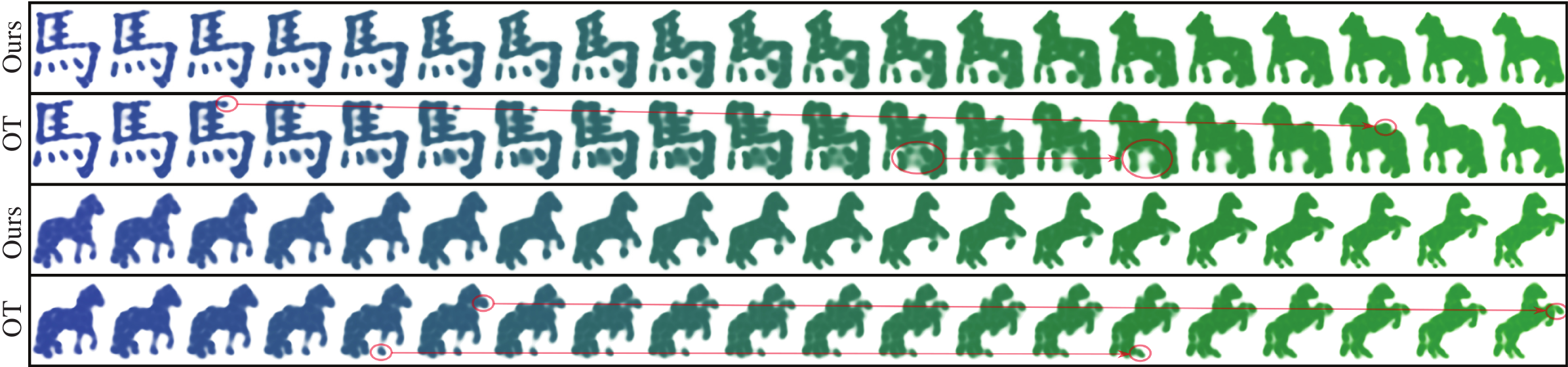}
    \vspace{-20pt}
    \caption{Comparison between trajectories obtained using our method and optimal transport for interpolating between the Chinese character for ``horse'' and an image of a horse (rows 1 and 2) and between two images of horses in different poses (rows 3 and 4). The OT interpolation \add{computed via \cite{flamary2021pot}} exhibits more spatial discontinuities (circled in red).}
    \label{fig:horsecharacter}
\end{figure*}

% \begin{figure*}
%     \centering
%     \includegraphics[width=\textwidth]{figures/horses_new.png}
%     \vspace{-20pt}
%     \caption{Comparison between trajectories obtained using our method and optimal transport for interpolating between the Chinese character for ``horse'' and an image of a horse (rows 1 and 2) and between two images of horses in different poses (rows 3 and 4). The OT interpolation exhibits more spatial discontinuities (circled in red).}
%     \label{fig:horsecharacter}
% \end{figure*}

\begin{figure}
    \centering
    %\includegraphics[width=1\columnwidth]{figures/rectdiv_withlabels.png}
    %\put(-250,50){\rotatebox{90}{\small \% Area increase}}
    %\put(-150,30){time}    
    %\includegraphics[width=1\columnwidth]{figures/bars_horiz.png}
    % \includegraphics[width=1\columnwidth]{figures/divandrig2.png}
    \includegraphics[width=1\columnwidth]{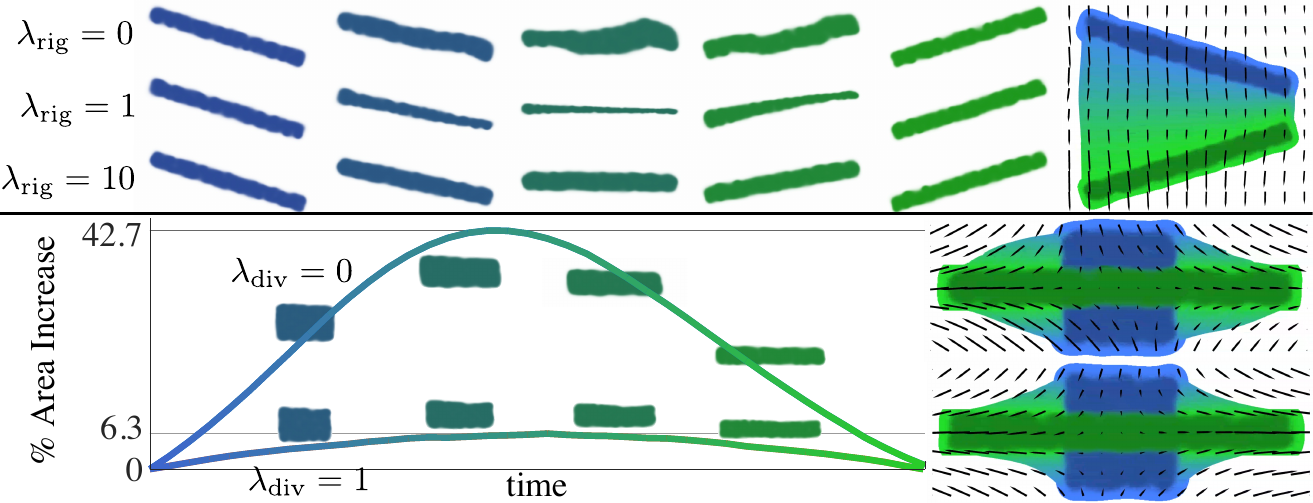}
    \vspace{-17pt}
    \caption{Effect of rigidity and compressibility regularization on our trajectories. Increasing the coefficient of the rigidity regularizer yields a less wobbly interpolation between two straight bars (top). Regularizing compressibility when interpolating between a square and rectangle reduces the increase in area of the shapes along the trajectory (bottom).}
    \label{fig:riganddiv}
\end{figure}

\begin{figure}
    \centering
\includegraphics[draft=false,width=1\columnwidth]{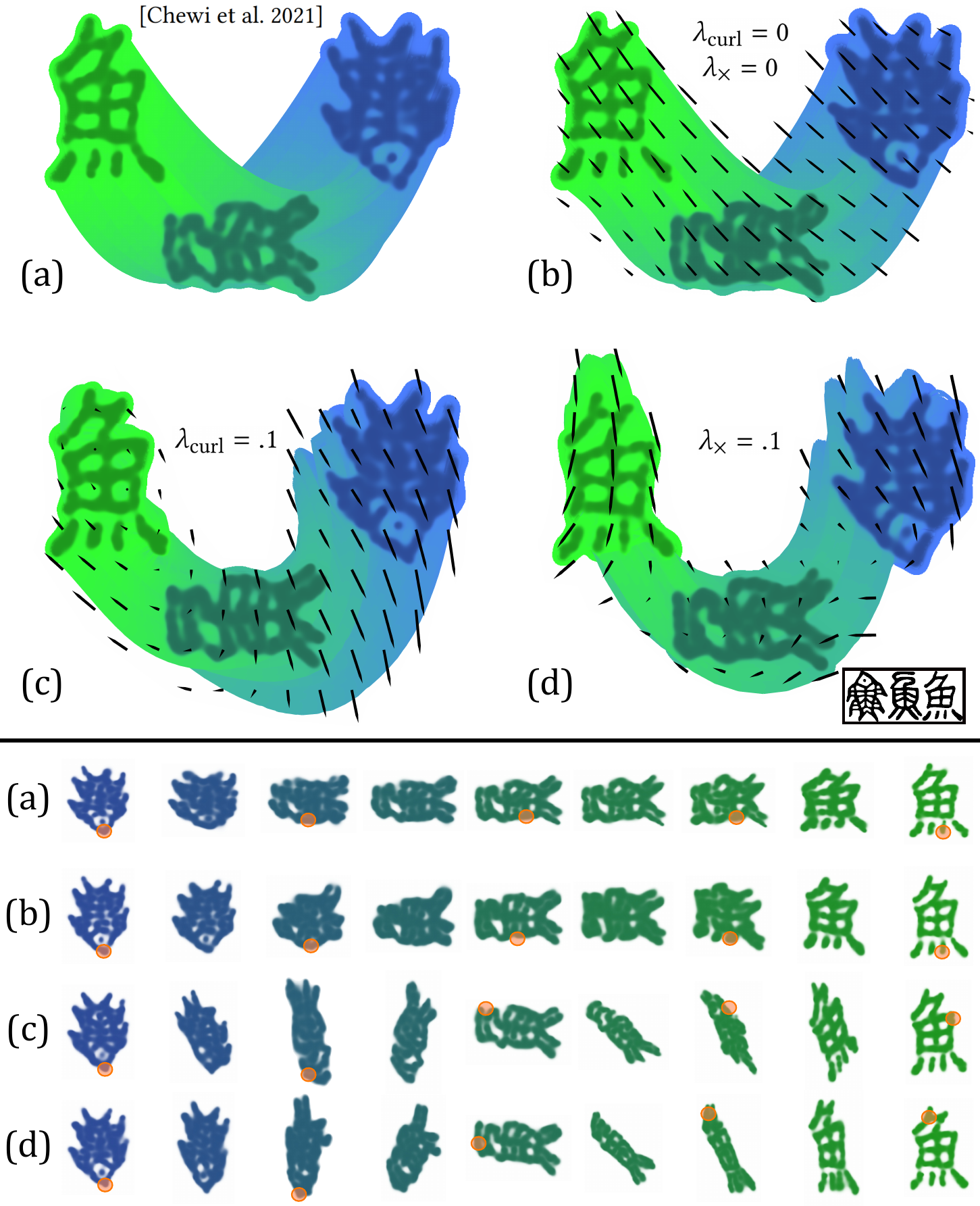}
    \vspace{-20pt}
    \caption{Interpolation between three keyframes, demonstrating the evolution of the Chinese character for ``fish.'' Without swirliness regularization ($\lambda_\lcurl=\lambda_\times=0$), the character does not properly rotate, with the head of the fish in the first keyframe (circled in orange) getting mapped to the tail in the last. Trajectories using \cite{chewi2021fast} exhibit the same issue. When we regularize the trajectory with $\lambda_\lcurl=10^{-1}$, the head of the fish in the first keyframe over-rotates, landing near the head of the final keyframe. When regularizing with $\lambda_\times=10^{-1}$, the head of the fish in the first keyframe is properly mapped to the head of the fish in the last keyframe. 
    }
    \label{fig:fishcurl}
\end{figure}

% \begin{figure*}
%     \centering
%     % \includegraphics[draft=false,width=1\textwidth]{figures/fish234_combined/curlfish_withcubic.png}
%     % \includegraphics[draft=false,width=1\textwidth]{figures/curlfish_simp.png}
%     % \includegraphics[draft=false,width=1\textwidth]{figures/curlfish2.png}
%     \includegraphics[draft=false,width=1\textwidth]{figures/fish234_trimmed.pdf}
%     \vspace{-20pt}
%     \caption{Interpolation between three keyframes, demonstrating the evolution of the Chinese character for ``fish.'' Without swirliness regularization ($\lambda_\lcurl=0$), the character does not properly rotate, with the head of the fish in the first keyframe (circled in orange) getting mapped to the tail in the last. Trajectories using \cite{chewi2021fast} exhibit a similar issue. We improve our trajectory by regularizing swirliness ($\lambda_\lcurl=10^{-1}$). }
%     \label{fig:fishcurl}
% \end{figure*}

\begin{figure}
    \centering
    \includegraphics[width=1\columnwidth]{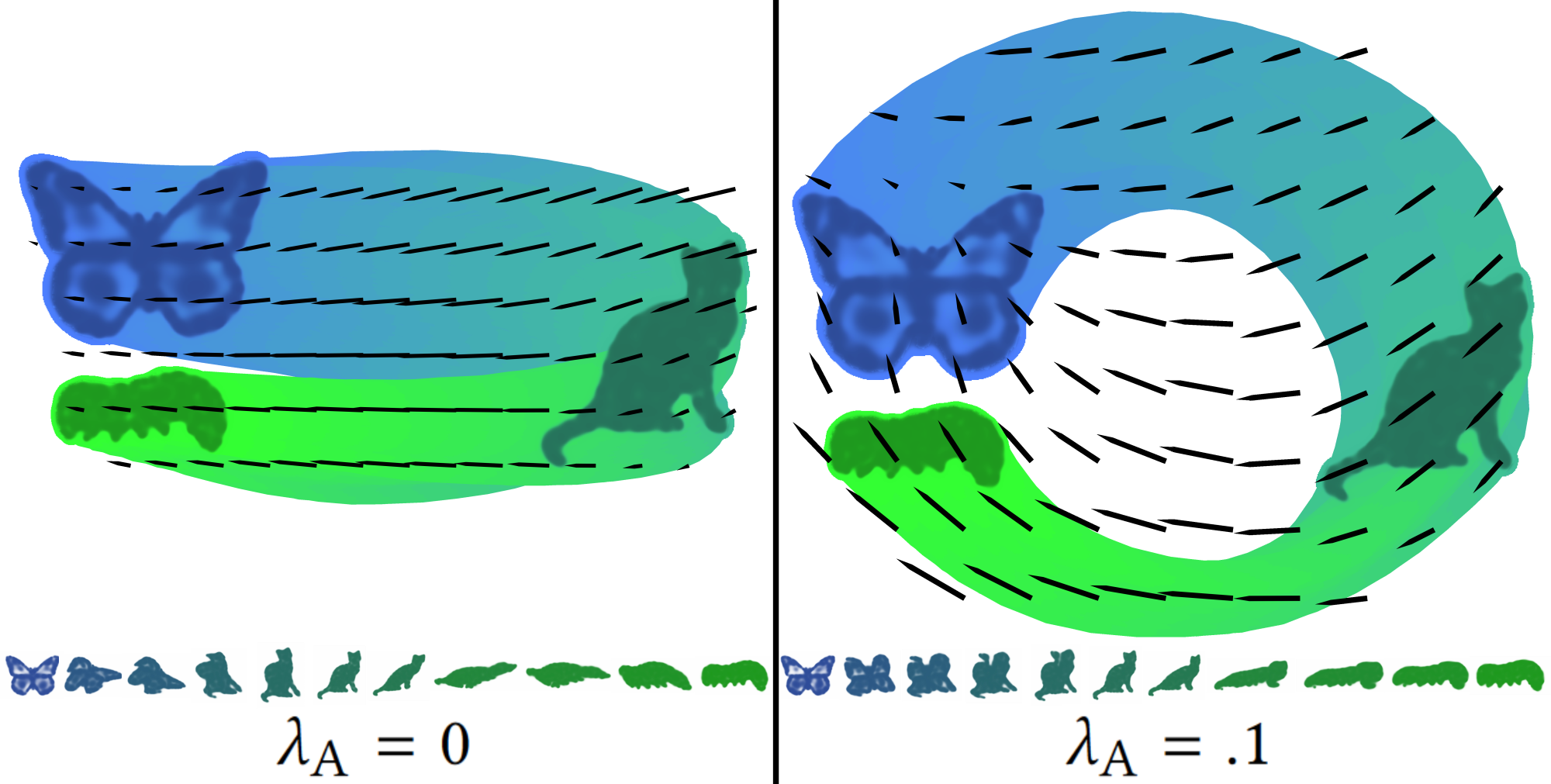}
    \vspace{-20pt}
    \caption{Effect of user-directed alignment on our trajectories. Adding this regularizer (right) to the original trajectory (left) yields a more circuitous path.}
    \label{fig:bcc_radial}
\end{figure}

\begin{figure}
    \centering
    \includegraphics[width=1\columnwidth]{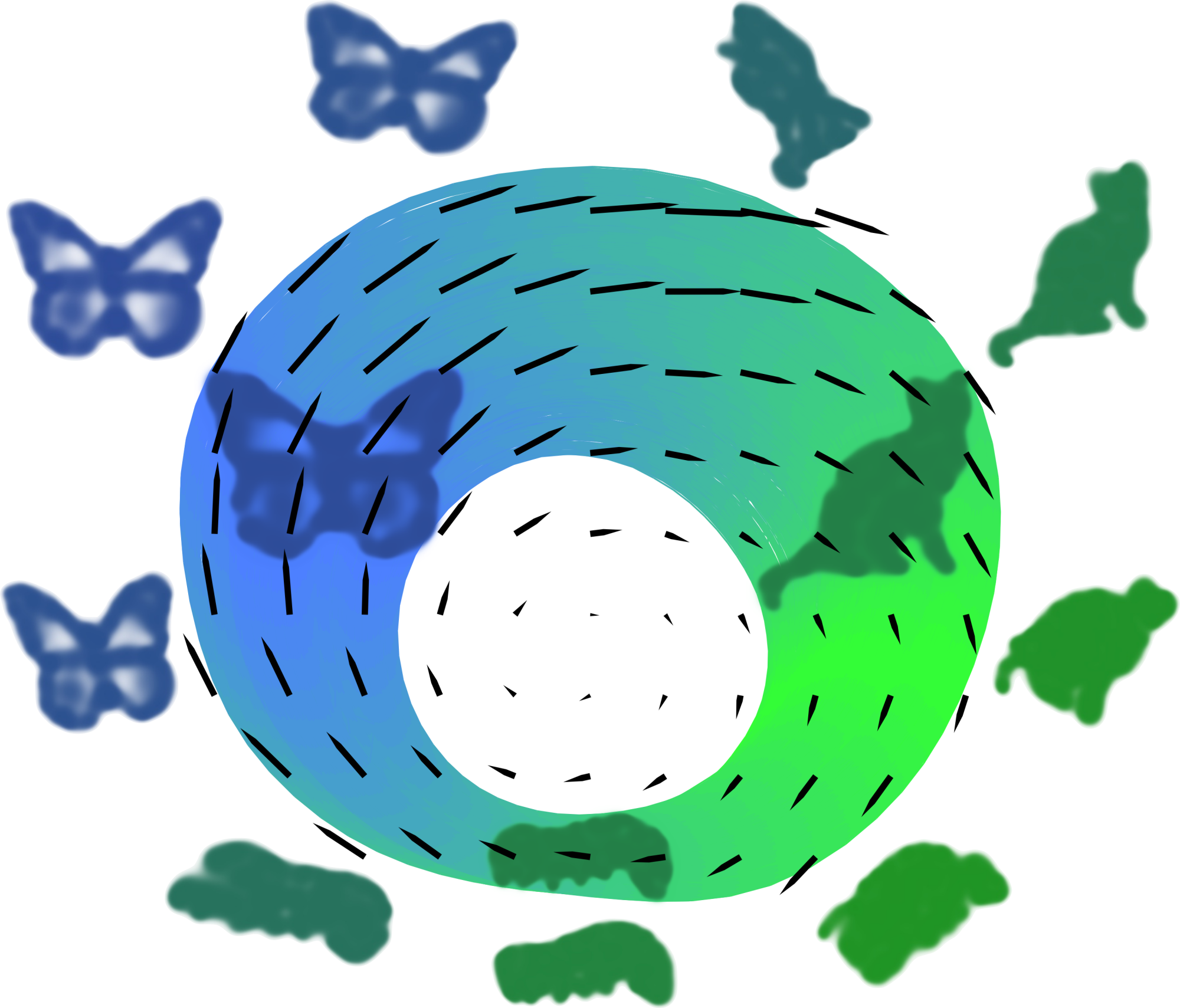}
    \vspace{-20pt}
    \caption{By combining user-directed alignment and carefully chosen Fourier frequencies we compute perfectly loopable circular trajectories depicting the fictitious lifecycle of a butterfly.}
    \label{fig:bcc_cyclic}
\end{figure}

% \begin{figure}
%     \centering
%     % \includegraphics[width=1\columnwidth]{figures/BCC_cycle/bcccycle.png}
%     % \includegraphics[width=1\columnwidth]{figures/BCC_base_vs_radial/bcc_bvr2.png}
%     % \includegraphics[width=1\columnwidth]{figures/bcc2.png}
%     \includegraphics[width=1\columnwidth]{figures/bcc_trimmed.pdf}
%     \vspace{-20pt}
%     \caption{Effect of user-directed alignment on our trajectories. Adding this regularizer (middle) to the original trajectory (left) yields a rounder path. This can be utilized to compute perfectly loopable circular trajectories (right).}
%     \label{fig:bcc_radial}
% \end{figure}

\begin{figure*}
    \centering
    \includegraphics[width=1\textwidth]{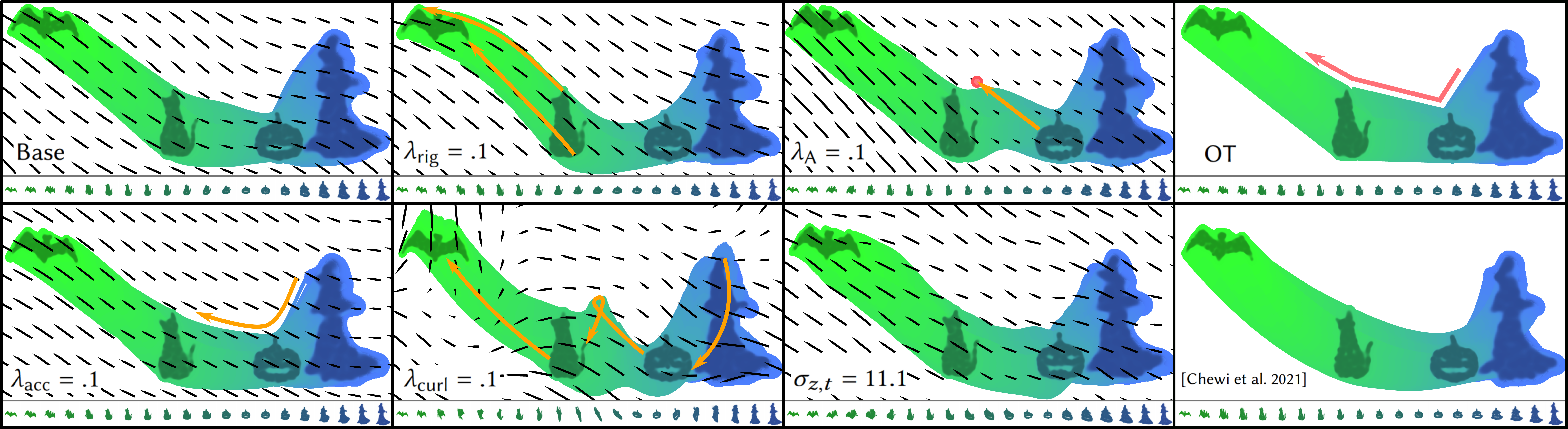}
    \vspace{-20pt}
    \caption{Effects of various regularizers on a four-keyframe animation. Employing rigidity, user-directed alignment, acceleration, and swirliness regularization as well as modifying the standard deviation of the RFF results in trajectories with varying path qualities and shapes. We also compare to trajectories obtained using \add{optimal transport \cite{flamary2021pot}} and \cite{chewi2021fast}.}
    \label{fig:hwnmed}
\end{figure*}

\subsection{Notation}

Let the keyframes be the set $\K = \{(\X_i, t_i)\}_{i=0}^{T-1}$, where each $\X_i$ is a measure over $\R^d$ for $d\in\{2,3\}$ from which we can draw samples; $t_i$ are corresponding timestamps. 
%We represent keyframes by point clouds $X_i$ and corresponding timestamps $t_i$ for $i \in \{0,...,t-1\}$. Each point cloud is embedded in dimension $d\in\{2,3\}$ and contains $n$ points i.e. $X_i \in \R^{n \times d}$. One could also consider a keyframe to be a measure over $\R^d$ formed by the sum of $n$ delta functions. 
% In practice, our point cloud keyframes are generated by sampling from indicator functions of sillhouettes.
Similarly to \citet{trajectorynet}, we compute trajectories between keyframes using a coordinate MLP 
\begin{equation}\label{eq:coordmlp}
f_\theta(z, t): \R^d \times \R \rightarrow \R^d
\end{equation}
and neural ODE $\dot{z}=f_\theta(z, t)$. This choice gives us easy access to all derivatives of $f_\theta$, something that would be more limited under a non-neural parameterization, e.g., if we parameterize $f_\theta$ on a grid with piecewise tri-linear basis functions, $\ddot{f_\theta}=0$.

For \add{times} $t_i \leq t_j$, let $\phi^{t_i,t_j}$ be the diffeomorphism of $\R^d$ resulting from integrating \eqref{eq:coordmlp} from time $t_i$ to time $t_j$. %corresponding map from $z(t_i)$ to $z(t_j)$. 
When $t_i > t_j$, $\phi^{t_i,t_j}$ is the inverse of $\phi^{t_j,t_i}$. 
Let $\X_i^{t_j}$ be \add{shorthand for} the pushforward of keyframe $\X_i$ by $\phi^{t_i,t_j}$:
\begin{equation}
    \X_i^{t_j} = \phi^{t_i,t_j}_{\# }\X_i.
\end{equation}
We sample from
% $\phi^{t_i,t_j}_{\# }\X_i$
$\X_i^{t_j}$
by evolving samples from $\X_i$ through the ODE \eqref{eq:ode} from time $t_i$ to time $t_j$.
In this notation, a trajectory strictly adhering to keyframes satisfies
\begin{equation}
    % \phi^{t_i,t_j}_{\# \X_i} = \X_j
    \X_i^{t_j} = \X_j.
    \label{eq:fitcond}
\end{equation}
%\justin{these X variables are weird to me.  Are they density functions?  Say what they are explicitly.}

\subsection{Fitting Loss}

CNF methods like \cite{trajectorynet,neuralode,ffjord} use Kullback-Leibler divergence $\KL(\X_j | \X_i^{t_j})$ as a fitting loss.
For  unstructured animation, however, a KL  loss is unsuitable because animation keyframes often have compact support; in particular, pushforward measures like $\X_i^{t_j}$ are unlikely to overlap with their targets $\X_j$ early in training. Furthermore, computing KL divergence requires density access, which is expensive to estimate. These situations leave the KL divergence undefined or infinite \cite{arjovsky2017wasserstein}. 

Instead, we use Sinkhorn divergence as a trajectory fitting loss: 
\begin{equation}
    L_{\lfit}^{i,j} = S_\epsilon(\X_i^{t_j} | \X_j).
    \label{eq:EfitIJ}
\end{equation}
A Sinkhorn divergence of exactly 0 guarantees that $\X_i^{t_j} = \X_j$. Unlike KL divergence, Sinkhorn divergence has no dependence on overlapping support between its measures. In addition, its gradient brings non-overlapping measures together. Finally, Sinkhorn divergence can be computed with only sample access from its input measures.

Our total trajectory fitting loss is then
\begin{equation}
    L_{\lfit} = \sqrt{\sum_{i=0}^{T-2} L_{\lfit}^{i,i+1} + L_{\lfit}^{i+1,i}}.
    \label{eq:Efit}
\end{equation}
% $L_{\lfit}$ is constructed through \autoref{eq:EfitIJ} applied to consecutive pairs of keyframes. 
This choice is motivated by ``teacher-forcing'' in training recurrent neural networks (RNNs), where one inserts ground-truth data into the network to decrease training time \cite{williams1989learning}. In our case, $L_{\lfit}$ is constructed by evaluating \autoref{eq:EfitIJ} only between consecutive keyframes, i.e., the ground truth data.
% This formulation is enabled by the Sinkhorn divergence, which only requires sample access. 
% A KL divergence--based loss would not be possible, since keyframe densities are not available. 
We also use a square root in \autoref{eq:Efit} %over the aggregated fitting energy. This is 
in preparation to balance our fitting loss against a regularization energy. Recall that 
$L_{\lfit}^{i,i+1}$ is essentially a squared Wasserstein distance and that gradients of squared quantities decrease with the magnitudes of their arguments. The square-root maintains the magnitude of the fitting loss gradient even as the fitting loss approaches 0. This ensures that fitting loss gradients are not out-competed by regularizers.

\subsection{Controlling the Trajectory}
\label{sec:regularizers}
% mention neural bias?
Within the space of trajectories that minimize the fitting loss, we can encourage paths with desirable features by regularizing the velocity field $f_\theta$. This process is straightforward, since types of motion have natural vector calculus or continuum mechanics counterparts. 
Here, we identify different types of motion with mathematical quantities based on vector field $f_\theta$ in a pointwise manner. These pointwise quantities will be denoted $l_\square$ with placeholder symbol $\square$.
We then show how to integrate these pointwise quantities in space-time to build a corresponding regularizing loss $L_\square$. 
% control the trajectory of an animation. 
These regularizers enable the user to select from a palette of options to control their animation.

\paragraph*{Squash and Stretch.} We begin with one of the basic principles of animation: ``squash and stretch'' \cite{thomas1995illusion}. A mathematical measure for this type of movement is \emph{rigidity}:
% linear strain tensor from continuum mechanics
\begin{equation}
    l_{\lrigid} = \|\nabla_z f_\theta + \nabla_z f_\theta^\top\|_F^2,
\end{equation}
where $\|\cdot\|_F$ denotes Frobenius norm. 
This term can also be interpreted as a Killing vector field energy \cite{ben2010discrete}, or, from the perspective of continuum mechanics, $f_\theta$ is the displacement gradient, and $l_\lrigid$ is the magnitude of the linear strain tensor. Movements generated by a velocity field where $l_\lrigid=0$ will appear rigid and not squash or stretch keyframes.

\paragraph*{Compressibility.} 
A closely related concept to rigidity is compressibility. 
Consider a quivering block of gelatin. While its movements are non-rigid, squashing in height will cause bulging in width. Compressibility can be captured by \emph{divergence}:
\begin{equation}
    l_{\ldiv} = (\nabla \cdot f_\theta)^2 = \Tr[\nabla_z f_\theta]^2.
\end{equation}
Movements of a keyframe generated by a velocity field where $l_\ldiv=0$ will preserve area or volume, i.e., they are incompressible.

\paragraph*{User-Directed Alignment.}
We can explicitly incentivize keyframes to move in certain directions during the animation via a user-provided metric $A(z,t)\in\R^{d\times d}$ by minimizing
%, where $A=I-vv^\top$ and loss
\begin{equation}
    l_{\lmetric} = \|f_\theta\|_A^2 = f_\theta^\top A f_\theta.
\end{equation}
If the user wants $f_\theta$ to align with unit vector field $v$, they can choose the metric $A = I-vv^\top$. Conversely, the user can also penalize alignment of $f_\theta$ to $v$ with $A=vv^\top$.

\paragraph*{Swirliness.} 
One of the most visually salient features of an animation or fluid is its swirliness. This is naturally captured by vector valued quantity \emph{curl}: $\nabla \times f_\theta$.
% Curl is vector-valued but in 2D, it is standard to treat it as a scalar. 
Its direction points in the axis of rotation, and its magnitude is twice the angular velocity. Given a target curl vector $\vec{c}$, we can quantify how close $f_\theta$ is to meeting that target with
\begin{equation}
    l_\lcurl = \|\nabla \times f_\theta - \vec{c}\|_2^2.
\end{equation}
For example, if we want a 2D animation where the keyframe rotates a full circle clockwise in four seconds, then we can measure $l_{\lcurl}$ with $\vec{c}=[0,0,-\pi]^\top$. 

$l_\lcurl$ allows us to quantify how much $f_\theta$ deviates from the target curl at any specific point in the trajectory. To quantify how much the integrated curl vector along a trajectory deviates from the target, we also need $l_\times = \nabla \times f_\theta$, the only vector valued $l_\square$ in our list of regularizers.

% Unlike the other quantities, $l_{\lcurl}$ is vector-valued but in 2D it is standard to treat it as a scalar.
% Its direction points in the axis of rotation, and its magnitude is twice the angular velocity. For example, if we have a 2D velocity field where $l_{\lcurl}+\pi = 0$, then keyframes following this trajectory will spin a full circle clockwise in four seconds. 

% Similarly, if one wanted trajectories with low rotation they might choose a curl-based regularizer
% \begin{equation}
%     l_{\lcurl} = \|\nabla_z f_\theta - \nabla_z f_\theta^\top\|_F^2.
% \end{equation}\justin{i see no curl in this formula. say more}
% Locally rigid trajectories can be obtained by interpreting $\nabla_z f_\theta$ as a displacement gradient, and minimizing the corresponding linear strain tensor
% \begin{equation}
%     l_{\lrigid} = \|\nabla_z f_\theta + \nabla_z f_\theta^\top\|_F^2.
% \end{equation}
% This can also be understood as incentivizing that $f_\theta$ be a Killing vector field.
% One can encourage more fluid like trajectories by using self-advection term 
% \begin{equation}
%     l_{\ladvect} = \| f_\theta \cdot \nabla_z f_\theta + \dot{f_\theta}\|_F^2.
% \end{equation}

\paragraph*{Smoothness.}
Finally, we mention some regularizers from the neural ODE literature. ODE solvers can be slow to converge if $f_\theta$ varies too much spatially or temporally \cite{kelly2020learning,finlay2020train}. One can try to make $f_\theta$ \emph{smoother} with the following:
% producing kinetic energy (KE), acceleration, and jerk based losses.
\begin{gather}
    l_{\lgrad} = \|\nabla_z f_\theta\|_F^2, \hfill \\
    l_{\lvel} = \|f_\theta\|_2^2,\;\;
    l_{\laccel} = \|\dot{f_\theta}\|_2^2, \;\;
    l_{\ljerk} = \|\ddot{f_\theta}\|_2^2.
\end{gather}
$l_{\lgrad}$ quantifies spatial variation of $f_\theta$, while $l_{\lvel}$, $l_{\laccel}$, and $l_{\ljerk}$ measure  orders of temporal variation, i.e., speed, acceleration, and jerk. 

\paragraph*{Integrating Along the Trajectory.}
We have defined various point-wise quantities based on $f_\theta$. We will denote all of them with $l_\square(z,t)$, where $\square$ can be replaced with any of the previously-mentioned quantities.
To regularize $f_\theta$ with $l_\square(z,t)$ we use the loss
\begin{equation}
    L_\square = \frac{1}{2} \sum_{i=0}^{T-2} 
    \int_{t_i}^{t_{i+1}} 
    % \frac{\int_{\R^d} l_\square(z,t) d\X_i^t + \int_{\R^d} l_\square(z,t) d\X_{i+1}^t}{2}
    \int_{\R^d} l_\square(z,t) d\left(\X_i^t + \X_{i+1}^t\right)
    dt
    \label{eq:Eregtraj}
\end{equation}
that integrates $l_\square$ over the trajectory dictated by $f_\theta$. 
As with \autoref{eq:Efit}, \autoref{eq:Eregtraj} only considers trajectories obtained by flowing keyframes to their consecutive neighbors. Doing this avoids applying regularization unnecessarily at space-time locations that come from accumulated error in ODE integration. 

We add regularizer $L_\square$ to our total loss with coefficient $\lambda_\square$.
% The only exception is $l_\lcurl$ in 3D which is a vector quantity, in which case the integrand is $\|l_\lcurl(z,t)\|_2^2$.
The one exception to this symbolic grouping ($\lambda_\square$, $L_\square$) is when $\square=\times$ because $L_\times$ is a vector. Let $\lambda_\times$ represent regularizing factor corresponding to regularizing loss function $\|L_\times - \vec{c}\|^2_2$ with target curl $\vec{c}$. $\lambda_\times$ regularizes the \emph{average} curl as opposed to $\lambda_\lcurl$ which regularizes \emph{pointwise} curl.
We show in \autoref{sec:results} the isolated effects of these regularizers and how they can be used to control an animation.
The total loss function is
\begin{equation}
    L_{\ltotal} = L_\lfit + \sum_{\text{scalars}} \lambda_\square L_\square + \lambda_\times \|L_\times - \vec{c}\|_2^2
\end{equation}

% Using \autoref{eq:Eregtraj} as a regularizer essentially minimizes  $l_\square$ uniformly over the trajectory. 

% Specifically for curl, we define one more loss term based on 
% \begin{equation}
%     L_\lct = \frac{1}{2} \sum_{i=0}^{T-2} 
%     \int_{t_i}^{t_{i+1}} 
% \int_{\R^d} (\nabla \times f_\theta) d\left(\X_i^t + \X_{i+1}^t\right)
%     dt
%     \label{eq:Eregtraj}
% \end{equation}

% One can also apply $l_\square(z,t)$ uniformly on a bounding box over keyframes $BB(\K)$.  
% \begin{equation}
%     L_\square^U = 
%     \int_{t_0}^{t_{T-1}} 
%     \int_{BB(\K)} l_\square(z,t) dz dt
%     \label{eq:Eregglob}
% \end{equation}
% This may be used to bring about effects such as global incompressibility.

% \justin{you'll need a figure showing the same trajectory with each of these turned on one at a time}
 
% cyclic via model modification
% thm about achieving dynamic OT?

\subsection{Wasserstein Barycenter Interpolation}
After training the neural ODE, we are not yet guaranteed that the trajectory strictly adheres to keyframes, as the neural ODE balances the fitting loss and regularizing losses. If $L_{\lfit}\neq0$, the neural ODE trajectories deviate from keyframes. 
%To ensure that our final trajectories strictly adhere to keyframes, 
We correct deviation by applying the following Wasserstein barycenter interpolation step. Given a query time $t \in [t_i, t_{i+1}]$, our output in-between frame is defined as
\begin{align}
    \vspace{-3pt}
    \X_{\epsilon,\tau}(t) = \argmin_\alpha  \left(t_{i+1} - t\right) & S_{\epsilon,\tau}(\X_i^t, \alpha) \nonumber
    \\
    + \left(t - t_i\right) & S_{\epsilon,\tau}(\alpha, \X_{i+1}^t).
    \label{eq:wassinterp}
    \vspace{-1pt}
\end{align}
In this way, our output trajectories are guaranteed to adhere to the keyframes: $\X_{\epsilon,\tau}(t_i)=\X_i$ for any choice of $\epsilon$ and $\tau$. 
This step of our pipeline is similar to \cite{solomon34convolutional}, who use Wasserstein barycenters for shape interpolation, but their barycenters are computed directly between keyframes while our barycenters are computed between ODE-advected keyframes allowing for artistic control.
When $L_{\lfit}=0$, \autoref{eq:wassinterp} becomes trivial with $\X(t) = \X_i^t = \X_{i+1}^t$. Our interpolation is illustrated schematically in \autoref{fig:wassinterp}.

%  The difference in result can be seen in \autoref{fig:horsecharacter}.
%\citet{solomon34convolutional} also compute Wasserstein barycenters for shape interpolation, however, their interpolation acts directly between keyframes while our interpolation acts on ODE advected keyframes. The difference in result can be seen in \autoref{fig:horsecharacter}.

\section{Implementation Details}
% \justin{wait a sec! i thought they were point clouds! i think what you mean to say is that the keyframes can be anything where you can sample point clouds, including...  maybe highlight that your method does \emph{not} need a density function, unlike some past CNFs?} 
% Our implementation, they are input as binary images for $d=2$, or triangle meshes for $d=3$. The densities of these keyframes are simply uniform over the binary image or the interior of the triangle mesh. From these densities it is easy to sample point clouds, by which we compute our various losses.

Here we describe various implementation details needed to replicate our results. The majority of figures are produced with the same default parameters though our method is not overly sensitive to these choices. 

Keyframes have been presented so far as measures over $\R^d$ with sample access. Our implementation mirrors this assumption and samples $N$ points per keyframe in every training iteration. For 2D (3D) examples, $N$ is initialized to $300$ $(1000)$ points.
We increase $N$ by a factor of $4^{\nicefrac{1}{6}} \sim 1.26$ every 50 training iterations, ensuring that $N$ has quadrupled by iteration 300.
All keyframes are jointly normalized to lie within $[-1,1]^d$. Our state derivative $f_\theta$ is a multilayer perceptron with normally sampled random Fourier features \cite{tancik2020fourier} of standard deviation $\sigma_{z,t}=\nicefrac{3 \pi}{\sqrt{d}} \sim 6.7$ in spatial and temporal dimensions. This distribution gives us fourier frequencies with periods roughly comparable to the diagonal of the bounding box.
We use three hidden layers of size 512 each. All nonlinearities are Tanh except for the final layer, which is Softplus. This choice of nonlinearities gives us easy access to all derivatives of $f_\theta$. In contrast, using all ReLU nonlinearities would have prevented us from effectively building regularizers on quantities like $\ddot{f_\theta}$.

We also employ incremental unmasking of 100 random Fourier features during training, as described in \citet{hertz2021sape}, with %the following modification. We unmask features at a rate 
a modified rate 
so that all features are unmasked when 80\% of training iterations are finished instead of their 50\%. All models were trained for a fixed 300 iterations with a learning rate of $10^{-4}$. We use Adam \cite{kingma2014adam} with default parameters and a learning rate scheduler that halves the learning rate on plateau with a minimum learning rate of $10^{-7}$.

The Sinkhorn divergence in \autoref{eq:EfitIJ} is computed via GeomLoss \cite{feydy2017optimal} with entropic regularization weight $\epsilon=10^{-4}$. 
Integration of the neural ODE is done using ``torchdiffeq'' with all default parameters \cite{neuralode}.
The space-time integral in \autoref{eq:Eregtraj} is computed each training iteration by sampling 30 points of keyframe $\X_i$, integrating in time to 5 uniformly sampled values in $[t_i,t_{i+1}]$, and averaging their $l_\square(z,t)$ values. We compute Wasserstein barycenters solving \autoref{eq:wassinterp} by initializing $\alpha=\X_i^t$ and iterating gradient descent via GeomLoss.

During training, we normalize $L_\lfit$ to start at $\sqrt{2}$ in the first iteration; unless stated otherwise, all results in \autoref{sec:results} are generated including $\lambda_\ljerk = 10^{-2}$ as a regularizer. This regularization ensures $f_\theta$ does not take convoluted trajectories that result in slow ODE integration. All input keyframes are one second apart. 
Our computations are performed on a single Nvidia GeForce RTX 3090 GPU and take approximately 10 (15) minutes to train a 2D (3D) animation. 

% \paragraph*{Partial gradients.}
% Computing gradients at training time can be expensive since it requires differentiating through an ODE solver. While this is unavoidable for gradient based optimization, we find that a shortcut can greatly accelerate early iterations of training. When computing gradients of \autoref{eq:Eregtraj}, we simply skip propagating gradients through the ODE solver. This amounts forgetting that the trajectories on which we aggregate regularizers are dependent on the neural weights $\theta$. Once the loss stops decreasing, we remove this shortcut and return to proper gradient estimation.

To visualize our trajectories, in 2D, we render each frame by splatting isotropic radial basis functions (RBFs) at each of 4000 point samples per frame. Following \citet{bonneel2011displacement}, we choose the bandwidth for each RBF as the distance to the 20th nearest neighbor of the corresponding point. All velocity fields in our figures correspond to the final time step. We also trace out the trajectories of point clouds through the animation to visualize the shape of the trajectory. This takes $\sim$1s to render per frame.
In 3D, we visualize frames using spherical metaballs, where the radius of each metaball is determined based on the corresponding point's distance to its 25th nearest neighbor. We smooth the resulting meshes using 20 iterations of the Blender ``Smooth'' modifier with a smoothing factor of 2. Point clouds for 3D renders contain 25000 points and take $\sim$10s per frame.

The metaball rendering for 3D can also be applied to our 2D results generally producing sharper boundaries, and lower interior variation as shown in \autoref{fig:doublerender}. Animations in \autoref{sec:results} are accompanied by videos in supplementary materials. The metaball rendering is also additionally applied to more 2D results in supplementary materials. We strongly recommend viewing the animations rather than relying solely on static figures within the paper.

%\subsection{Renders}
%\cite{bonneel2011displacement} RBF KNN 20
\section{Results}
\label{sec:results}
\paragraph*{Optimal Transport.}
We compare trajectories obtained using optimal transport and our method in \autoref{fig:horsecharacter}. In the first two rows, we interpolate from the Chinese character for ``horse'' to an image of a horse, and in the second two rows we interpolate between two images of horses in different poses. In both cases, the OT interpolation has more spatial discontinuities (circled and tracked in red). Since our method constructs a diffeomorphism by integrating \autoref{eq:coordmlp}, our trajectories tend to avoid discontinuous movements. % Mass splitting occurs in our trajectories only due to \autoref{eq:wassinterp} but is much less apparent than direct OT.
Our results generate more intuitive interpolations between keyframes than OT, which maps points at the top of the horse character to its back.

\paragraph*{Rigidity and Compressibility.}
\autoref{fig:riganddiv} tests the effect of $L_\lrigid$ and $L_\ldiv$ on our trajectories. 
% All trajectories in this figure include $.1 L_\lvel$ and no $L_\ljerk$.
In the top half, we interpolate from the blue bar to the green bar with $\lambda_\lvel = 10^{-2}$, $\lambda_\ljerk = 0$ and
$\lambda_\lrigid = \{0, 1, 10\}$. The $\lambda_\lrigid=0$ trajectory is wobbly and bends the bar severely. The $\lambda_\lrigid=1$ trajectory is straighter but still shows some deformation. The $\lambda_\lrigid=10$ case is almost entirely rigid. 
In the bottom half, we interpolate from a blue unit square to a green rectangle with width 3 and height $\nicefrac{1}{3}$. The plot shows percent area increase throughout the interpolation with $\lambda_\ldiv=0$ vs.\ $\lambda_\ldiv=1$. In both cases, we also include $\lambda_\lvel = 10^{-1}$ and $\lambda_\ljerk=0$. When $\lambda_\ldiv=0$, the trajectory increases area by 42.7\%, and when $ \lambda_\ldiv=1$, the area increase drops to 6.3\%. On the right, we show the keyframes and their trajectories traced out for $\lambda_\ldiv=10^{-1}$. The trajectories are qualitatively different, e.g., when $\lambda_\ldiv=0$, the trajectory traces out a concave path, but when $\lambda_\ldiv=1$, the trajectory traces out a convex path with more area preservation.

\paragraph*{Curl Regularization.}
\autoref{fig:fishcurl} interpolates between three keyframes from the etymology of the Chinese characters for ``fish.'' These characters are shown upright in the black box at the right of the figure. As pictographs, they are meant to be similar to fish, with their bottoms resembling tail fins and tops resembling fish heads. For the animation, we lay these characters in a semicircular arrangement, with the intention that an animated trajectory between them should follow a semicircular arc. 

First, we show trajectories computed using  \cite{chewi2021fast} and mark in the orange circles corresponding points throughout the animations. The markers show that the head of the fish in keyframe 1 is mapped to the tail in keyframe 3. Since their method computes OT maps between consecutive keyframes, and the vector fields producing OT interpolations are necessarily curl free, this result is unsurprising. 
Then, we compute trajectories using the default $\lambda_\ljerk=10^{-2}$. Again, the intermediate fish characters do not rotate through the animation. 
In the bottom left, we show the effect of adding $\lambda_\lcurl=10^{-1}$  with a target pointwise curl of $\vec{c}=[0,0,-\pi]^\top$. As described in \autoref{sec:regularizers}, this encourages the pointwise curl of $f_\theta$ to be $\vec{c}$ throughout the trajectory, corresponding to a clockwise rotation at the rate of $\nicefrac{\pi}{2}$ per second. The resulting animation is significantly different from before. The head in keyframe 1 corresponds to the head in keyframe 2, and the trajectory slightly over-rotates just past the head of the fish in keyframe 3; 
in the bottom right, we regularize with $\lambda_\times=10^{-1}$ incentivizing the average curl over the trajectory to be $\vec{c}$. In this case, the head in keyframe 1 is successfully mapped to the head in all following keyframes.
% $\lambda_\lcurl$ and $\lambda_\times$ both increased the swirliness of the animation.

\paragraph*{User-Directed Alignment and Cyclic Trajectories.}
\autoref{fig:bcc_radial} demonstrates the effect of $L_\lmetric$. We interpolate from a butterfly to a cat and finally to a caterpillar. The baseline trajectory is shown on the left. On the right, we add a $\lambda_\lmetric=10^{-1}$ regularizer that penalizes alignment of the velocity field to the unit radial vector field. As a result, the trajectory takes a \delete{round} circuitous path. 

In \autoref{fig:bcc_cyclic}, we compute a cyclic trajectory by repeating the first keyframe at the end of the keyframe list. We round temporal RFF coefficients to the nearest values that produce cyclic signals with a 3 second period. Finally, we add the same $\lambda_\lmetric=10^{-1}$ regularizer as in the middle to encourage a circular trajectory. The result is a perfectly loopable animation depicting the fictitious life cycle of a butterfly.

\paragraph*{Regularizing Trajectories.}
\autoref{fig:hwnmed} demonstrates the isolated effects of various regularizers on the trajectory of a four-keyframe animation. The goal is to transform from a witch, into a pumpkin, into a cat, and finally into a bat. By employing different regularizers, we achieve varying effects on the animation. 

In the top left, we show a baseline trajectory with just $\lambda_\ljerk=10^{-2}$ as a regularizer. Here, in-between frames maintain a mostly upright posture, i.e., the top of each keyframe is mapped to the top of the following keyframe. When the aspect ratios of the keyframes differ, this trajectory squashes keyframes into one another. 

Next, we impose an additional $\lambda_\lrigid=10^{-1}$. While there is no truly rigid interpolation between these keyframes, the cat rotates sideways into the bat, yielding a more rigid trajectory than the base case. The orange arrows indicate the path from the cat to the bat.

We then replace the rigid regularizer with $\lambda_\lmetric=10^{-1}$, where the metric is chosen to incentivize alignment of $f_\theta$ to the unit radial vector field. The origin is plotted in red. Due to $\lambda_\lmetric$, the trajectory of the pumpkin to the cat is pulled towards the origin, creating a bouncing effect.

As a comparison, on the top right, we show the trajectory from concatenated OT maps \add{computed by \cite{flamary2021pot}}. This results in extremely sharp turns at the keyframes, which is expected since trajectories are built piecewise.

On the bottom left, we impose $\lambda_\laccel=10^{-1}$ on top of the base. The shape of the trajectory at the pumpkin keyframe is smoother compared to the trajectory taken in the base case. 

Next we replace $\lambda_\laccel$ with $\lambda_\lcurl=10^{-1}$, where again, the curl is incentivized to be $\vec{c}=[0,0,-\pi]^\top$. As a result, the trajectory makes almost a full $2\pi$ rotation. 

Then we remove all regularizers and increase the RFF standard deviation from $\sigma_{z,t}=6.7$ to $11.1$. The increased magnitude of RFF coefficients and lack of regularization yield a much noisier trajectory.

Finally on the bottom right, we show \cite{chewi2021fast} for contrast. It produces a smooth path of cubic splines that are qualitatively close to the $\lambda_\laccel=10^{-1}$ case. Similar to \autoref{fig:fishcurl}, the trajectory does not rotate keyframes, instead opting to squash them to get the right aspect ratio.

\begin{figure}
    \centering
    \includegraphics[width=\columnwidth]{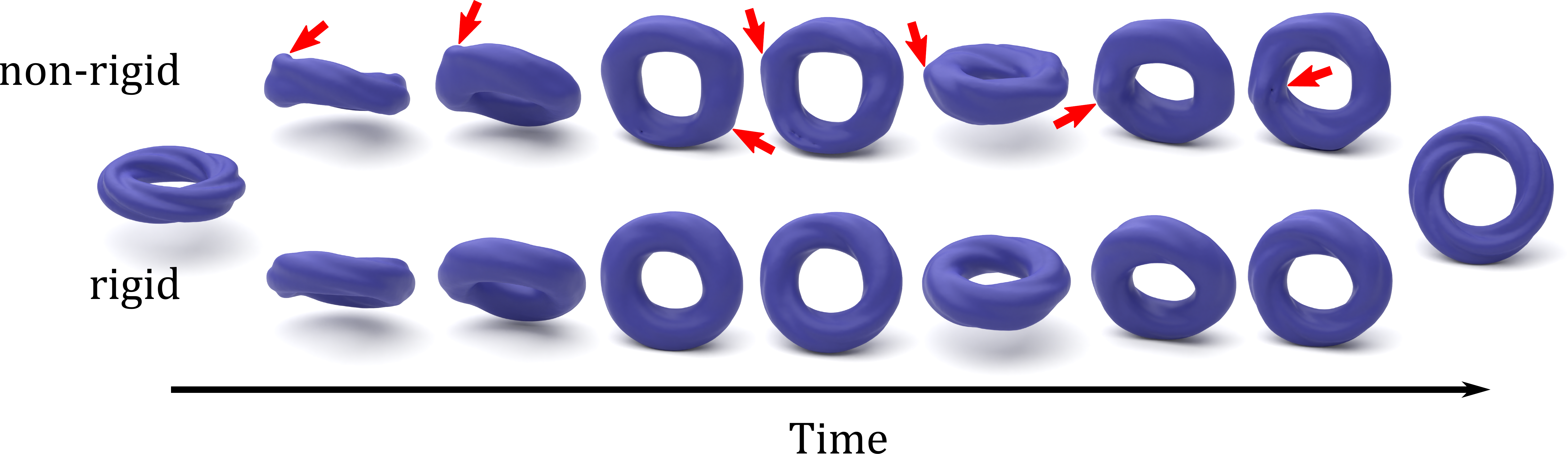}
    \vspace{-15pt}
    \caption{Effect of rigidity regularizer on trajectories of tori. 
    The top row is computed with default parameters, while the bottom row is computed with additional regularization $\lambda_\laccel=\lambda_\lrigid=10^{-1}$. Red arrows point out several locations where extra deformation occurs without rigidity regularization. Using these regularizers results in less wobbling along the trajectory. }
    \label{fig:rigidrings}
\end{figure}

\begin{figure}
    \centering
    \includegraphics[width=\columnwidth]{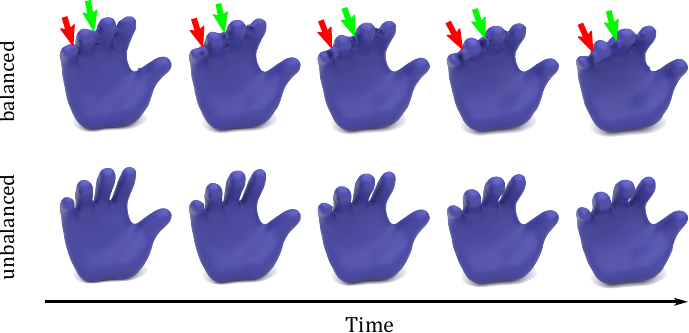}
    \vspace{-15pt}
    \caption{Effect of balanced v.s. unbalanced Wasserstein barycenter interpolation on intermediate frames of a hand closing animation. Volume of fingers is not carefully tuned in keyframes, resulting in balanced OT necessarily transporting mass between fingers during the animation. This is pointed out in the balanced interpolation, where red and green arrows track lumps of mass moving between fingers. Using unbalanced Wasserstein barycenters instead alleviates the mass splitting, and fingers are clearly separated throughout the animation. }
    \label{fig:balanced_hand}
\end{figure}

\paragraph*{Volumetric Results.}
In \autoref{fig:rigidrings}, we show various frames of an animation of a spinning ring with and without regularizers. In the first row we only use the default regularizer $\lambda_\ljerk=10^{-1}$. Since the loss does not care about rigidity, the resulting trajectory develops bulges indicated by the red arrows. In the second row we additionally regularize with $\lambda_\laccel=\lambda_\lrigid=10^{-1}$ to produce a much tamer trajectory. 

In \autoref{fig:balanced_hand}, we show the effect of unbalanced Wasserstein barycenter interpolation on interpolations of the same keyframes and the same neural ODE. The only difference is in the Wasserstein barycenter interpolation step \autoref{eq:wassinterp}, where the first row is computed with 
default parameter $\tau=\infty$,
and the second row with 
$\tau=0.05$.
Recall from \autoref{sec:prelims_unbalanced} that $\tau$ represents how much distance a transport plan is willing to move mass before giving up on constraints of the classical OT problem. Since fingers of the hand animation keyframes have different volumes, animations generated with $\tau=\infty$ necessarily transfer mass between fingers. This is tracked by the red and green arrows indicating mass moving between fingers. When we use unbalanced Wasserstein barycenters, the mass balancing between fingers is alleviated resulting in much sharper distinctions between each individual finger.

\autoref{fig:3d-traj} summarizes application of our method to generate 3D animations. In the first row, we build in-between frames for an animation from an open hand, to a partially closed hand, and finally to a cat. This animation is regularized with $\lambda_\laccel=\lambda_\lrigid=10^{-1}$. We use unbalanced Wasserstein barycenter interpolation with $\tau=.05$ for in-between frames from the open hand to the partially closed hand. Due to the increased complexity of computing unbalanced Wasserstein barycenters, we reserve the unbalanced case for only where we explicitly want to alleviate mass splitting and otherwise default to balanced barycenters.
In the second row, we interpolate from a sphere to a cow to a torus with all default parameters. \add{The change in topology is handled seamlessly.} In the last row, we interpolate through five keyframes of rings at different angles. The first, third, and fifth keyframes are rings with detailed helical patterns carved into them, while the second and fourth keyframes are normal tori. We regularize this animation with $\lambda_\laccel=\lambda_\lrigid=10^{-1}$ to produce a smooth trajectory.

% \begin{figure}
%     \centering
%     \includegraphics[draft=false,width=1\columnwidth]{figures/HWN/HNW_OT_cubic.png}
%     \caption{hwn}
%     \label{fig:HwnOTAndCubic}
% \end{figure}

% \begin{figure}
%     \centering
%     \includegraphics[width=1\columnwidth]{figures/BCC_cycle/bcc_new.png}
%     % \includegraphics[width=1\columnwidth]{figures/BCC_cycle/bcccycle.png}
%     % \includegraphics[width=1\columnwidth]{figures/BCC_cycle/bcccycle_noborder.png}
%     \caption{bcccycle}
%     \label{fig:bcccycle}
% \end{figure}

\section{Discussion and Conclusion}
% \FloatBarrier % this just moved text down instead of moving figure up :(
% mention unbalanced OT maybe
Unstructured animations appear in various forms of media ranging from hand-drawn to video games and film. These animations share a fluid morphing capability that is mesmerizing to watch but challenging to construct. Our work identifies unstructured animation as a density interpolation problem and builds automatic solutions through the machinery of optimal transport, neural ODEs, and PDE-based regularizers for intuitive and varied control.  

Future work might consider discontinuous parameterizations of the velocity field. Depending on the context, spatially discontinuous trajectories may be desirable, but integration of a smooth velocity field will always produce a diffeomorphism. Discontinuous velocity fields provide added flexibility, but also pose challenges to gradient based optimization and ODE integration.
A natural regularizer in this setting is vectorial total variation \cite{goldluecke2010approach}, which measures vector field smoothness but does not diverge near discontinuities.

Another avenue for further exploration might be %towards multi-scale parameterizations of the velocity field that can dramatically improve runtime. 
% Another way to reduce runtime might be 
to treat the fitting loss as a constraint. If the fitting loss can reach exactly 0, Wasserstein barycenter interpolation would no longer be necessary. The final rendering quality of our animations depend in part on the number of samples used to build the trajectory. Since the Wasserstein barycenter interpolation is computed independently per in-between frame and scales in expense with the number of points, it would be ideal to skip computing barycenters altogether.

% We expect that there are challenging theoretical questions as well about convexity of the optimization when using particular regularizers such as a user prescribed space-time metric. The neural parameterization obfuscates any convexity in the problem but $f_\theta$ is primarily a convenient discretization choice. 

Mesh- and rig-based animation are approachable for beginners through the abundance of accessible tools and tutorials. Unstructured animation is the opposite: Almost no documented computational tools exist enabling its design. This paper represents a step toward bridging the gap between rig-based animation and unstructured animation. We hope that the graphics community will discover more exciting approaches toward unstructured animation to further improve its ease of construction and accessibility.

\section*{Acknowledgements} 

The MIT Geometric Data Processing group acknowledges the generous support of Army Research Office grants W911NF2010168 and W911NF2110293, of Air Force Office of Scientific Research award FA9550-19-1-031, of National Science Foundation grants IIS-1838071 and CHS-1955697, from the CSAIL Systems that Learn program, from the MIT--IBM Watson AI Laboratory, from the Toyota–CSAIL Joint Research Center, from a gift from Adobe Systems, from an MIT.nano Immersion Lab/NCSOFT Gaming Program seed grant, and from a Google Research Scholar award.  This work was also supported by the National Science Foundation Graduate Research Fellowship under Grant No. 1122374. Paul Zhang acknowledges the support of the Department of Energy Computer Science Graduate Fellowship and the Mathworks Fellowship. The authors thank Aude Genevay and Christopher Scarvelis for many valuable discussions.

\printbibliography
\end{document}